\documentclass[prl,10pt,a4paper,nofootinbib]{revtex4-1}

\usepackage{color}
\definecolor{blue}{rgb}{0,0,0} %% -- BLAU
\definecolor{red}{rgb}{0,0,0}
\definecolor{soeren}{rgb}{0,0,0} %% -- SCHWARZ
\newcommand{\PANDA}{$\overline{\textrm{P}}\textrm{ANDA}$}
\usepackage{lineno}
\begin{document}

\thispagestyle{empty}

\huge

\begin{center}

\vspace*{4cm}

Input of the Belle II Collaboration \\

to the NuPECC Long Range Plan 2024

\vspace*{2cm}

\Large

Authors: B.\ G.\ Fulsom, J.\ S.\ Lange\footnote{Email {\tt soeren.lange$@$exp2.physik.uni-giessen.de}}, J.\ Libby,
E.\ Prencipe\footnote{Email {\tt elisabetta.prencipe$@$gmail.com}}, R.\ Seidl, C.\ P.\ Shen, A.\ J.\ Schwartz, B.\ A.\ Shwartz, D.\ Tonelli, A.\ Vossen

\vspace*{1cm}

{\bf Topic:} Hadron Physics 

\end{center}

\vspace*{1cm}

\Large 

{\bf Abstract:} The Belle II physics potential for hadron spectroscopy, inputs to $g$-2, fragmentation functions,
proton form factors, measurements of $\alpha_S$ and hadronic $\tau$ decays is outlined. 

\begin{center}

\vspace*{4cm}

September 30, 2022 

\end{center}

\newpage  

\setcounter{page}{1}

\normalsize

Possible contributions of the Belle II experiment to potential physics questions
in the NuPECC Long Range Plan (LRP) 2024 will be discussed. 
We outline the Belle II physics potential for hadron spectroscopy, inputs to $g$-2, fragmentation functions,
proton form factors, measurements of $\alpha_S$ and hadronic $\tau$ decays.
For additional details we refer to \cite{snowmass_belle2}.

\vspace*{-0.7cm}

%%%%%%%%%%%%%%%%%%%%%%%%%%%%
\section{Status of Belle II}
%%%%%%%%%%%%%%%%%%%%%%%%%%%%

\vspace*{-0.3cm}

Belle II is an hermetic spectrometer surrounded by particle-identification (PID) detectors, an electromagnetic calorimeter, and muon detectors
which reconstruct the final state particles of 7-on-4 GeV $e^+e^-$ collisions at the highest luminosities ever achieved, produced by the SuperKEKB collider
at KEK, Tsukuba, Japan. The primary scientific objectives are: indirect searches for non-standard-model physics using the weak interaction of quarks,
direct searches for low-mass dark matter, and studies of low-energy quantum chromodynamics (QCD) using spectroscopy.
At the time of writing this document, Belle II has recorded data sets corresponding to an integrated luminosity of 424~fb$^{-1}$.
After optics commissioning and background studies from February to June 2016,
a pilot run without vertex detectors, commissioning of the positron damping ring and first collisions were performed from April to July 2018.
Physics data taking started in 2019, with a one-layer silicon pixel detector (PXD) installed.
In July 2022 a long shutdown commenced for installation of a two-layer PXD. 
A peak luminosity of $4.7 \times 10^{34}$~cm$^{-2}$~s$^{-1}$ was reached. This new world record is about a factor of two higher compared to Belle. 
An integrated luminosity corresponding to 362~fb$^{-1}$ was recorded at the $\Upsilon$(4S) resonance, which decays almost 99\% to $B$ mesons\footnote{With a cross section of 1.09~$nb$,
an integrated luminosity of 1~fb$^{-1}$ corresponds to $1.09 \times 10^6$ produced $B^0 \overline{B^0}$ or $B^+ B^-$.}. 
For a comparison, BaBar and Belle recorded 424~fb$^{-1}$ and 711~fb$^{-1}$ at the $\Upsilon$(4S), respectively.
A data set of 42~fb$^{-1}$ was recorded in the continuum, $e.g.$ 60 MeV below the $\Upsilon$(4S).
Belle II is planned to reach a peak luminosity of 6$\times$10$^{35}$~cm$^{-2}$~s$^{-1}$, and collect an integrated luminosity of 50~ab$^{-1}$ in about 10 years of data taking. 

One of the improvements of Belle II compared to Belle is the redesign of the interaction region. 
The size of the overlap region of the colliding beam bunches
was decreased from 100~$\mu$m (horizontal) $\times$ 2~$\mu$m (vertical) to 10~$\mu$m (horizontal) $\times$ 59~nm (vertical).
This corresponds to the worlds smallest vertical $\beta$ function
where the  Gaussian width of a single beam near the IP is proportional to the square root
of a so-called beta function $\sigma_y = \sqrt{e \cdot \beta_y}$, $e$ = emittance.
Belle II shows so far the smallest vertical $\beta$ function worldwide. 
Due to its small size this is often referred to as the {\it nanobeam} scheme. Several subdetectors were upgraded for Belle II. 
Charged tracks in the redesigned central drift chamber are about 28\% longer than at Belle. 
New readout electronics in the electromagnetic calorimeter provides waveform sampling of 2~MHz for pulse shape discrimination. 
Particle identification is performed with a new Cherenkov detector based upon time-of-propagation in the radiator with about 40~ps timing resolution. 
The new 6-layer silicon vertex detector provides an impact resolution of about 25~$\mu$m in beam direction, which is about a factor of two improvement compared to Belle.

\vspace*{-0.8cm}

%%%%%%%%%%%%%%%%%%%%%%%%%%%%%
\section{Hadron spectroscopy}
%%%%%%%%%%%%%%%%%%%%%%%%%%%%%

\vspace*{-0.3cm}

%%%%%%%%%%%%%%%%%%%%%%%%%%
\subsection{Exotic mesons}
%%%%%%%%%%%%%%%%%%%%%%%%%%

\vspace*{-0.3cm}

In the last two decades, a number of narrow, exotic states have been observed, often denoted as $XYZ$ states \cite{xyz_review}.
They have been already  subject of interest in the LRP 2017. 
In the next decade, Belle II will be able to contribute significantly to their spectroscopy. 
Expected yields in a data set of 50~ab$^{-1}$ at Belle II are listed in Tab.~\ref{tyield}.
While these numbers may still be considered limited from the point of view of statistics, compared to yields at hadron colliders, they are particular notable strengths of Belle II. 

\begin{table}[hhh]
  \begin{center}
  \begin{tabular}{|l|l|l|}
    \hline
    State & Production and decay & N \\
    \hline
    \hline
    $X$(3872) & $B \rightarrow K X$(3872), $X$(3872)$\rightarrow J/\Psi \pi^+ \pi^-$ & $\simeq$14,000 \\
    $Y$(4260) & ISR, $Y$(4260)$\rightarrow J/\Psi \pi^+ \pi^-$ & $\simeq$29,600 \\ 
    $Z$(4430) & $B \rightarrow K^{\mp} Z$(4430), $Z$(4430)$\rightarrow J/\Psi \pi^{\pm}$ & $\simeq$10,200 \\
    \hline    
  \end{tabular}
  \end{center}
  \caption{Extrapolated yield $N$ for selected $XYZ$ states on 50~ab$^{-1}$ at Belle II, based upon the measured yields at Belle.\label{tyield}}
  \label{tyield}
\end{table}

\begin{itemize}

\item {\color{blue} The narrow total width of the $X$(3872) with ${\cal O}$(1~MeV) is remarkable,
  as it is about two orders of magnitude smaller than potential model predictions
  for the $\chi_{c1}'$, a predicted and yet unobserved charmonium state with nearby mass and identical quantum numbers.
  As the width is related to the wavefunction, 
  its precise knowledge could constrain theoretical models and help understanding the nature of the $X$(3872).} 
  All former width measurements have used the  $X(3872) \rightarrow J/\psi \pi^+ \pi^-$ decay.   At Belle II a new method will make use of the  $X(3872) \rightarrow D D \pi$ decay, having  a small Q value of only 7~MeV.  Preliminary studies show that the X(3872) width can be measured down to 950 (280)~keV with 3$\sigma$ significance for 10 (50)~ab$^{-1}$ \cite{belle2_width_x3872}, 
  and therefore reach sub-MeV precision. 
  While this is still larger than the expected measurable width of 40$\pm$13~keV by \PANDA, which  makes  use of a complementary technique of a resonance scan \cite{panda_width_x3872},   it is however much smaller than the Breit-Wigner width of 1.39$\pm$0.24$\pm$0.10~MeV measured by LHCb \cite{lhcb_width_x3872}. 
  Investigation of the lineshape by Belle II is presently under discussion. 
  
\item $e^+ e^-$ machines are unique to investigate ISR (initial state radiation) processes.
  In such a process, a photon is radiated in the initial state,
  and this way provides a way to measure a cross section while effectively changing the center-of-mass (c.m.)\ energy. 
  A number of $Y$ states in the charmonium region have been observed in ISR. {\color{blue} Those are far above the open charm threshold, but their decays to $D^{(*)} \overline{D}^{(*)}$ are strongly suppressed.
    Therefore they have been discussed as possible hybrid states $[c \overline{c} g]$ with a valence gluon.
    Only BESIII and Belle II can investigate these states.} 
  
\item $Z$ states are charged and therefore they are inherently exotic:
  in the charmonium ($Z_c$) and bottomonium ($Z_b$) mass region this implies a minimum quark content of four quarks. 
  There are charged $Z$ states in the charmonium region which are observed in $B$ meson decays.   They seem not related to any thresholds and have large widths in the order of $10^2$~MeV.   However, different $Z$ states have been observed in $e^+ e^-$ direct production or ISR, which are close to thresholds and have narrower widths in the order of 10~MeV.
  Belle II is the only experiment worldwide able to investigate both production processes with the same detector, and address the question of  why there are those two distinctive classes.

\item {\color{soeren} SuperKEKB can operate at c.m.\ energies above $\Upsilon$(4S) up to $\simeq11$~GeV,  including the $\Upsilon$(5S) and $\Upsilon$(6S) resonances,   which allow to investigate $Z_b$ states.
  As an example, Belle II has already recorded a data set of 19~fb$^{-1}$ around 10.75 GeV for the investigation   of the newly discovered, and possibly exotic,  $\Upsilon$(10753) structure.   These states have not been observed  at any hadron collider.}
  It also remains an interesting question that the partial decay width of $\Upsilon(5S) \rightarrow \Upsilon$(nS) $\pi^+ \pi^-$ is 
  higher than that of $\Upsilon (4S) \rightarrow \Upsilon$(nS) $\pi^+ \pi^-$ by an enormous factor of $\simeq$10$^3$ \cite{belle_y5s_high}. 
  
\item {\color{blue} BESIII \cite{bes3_x3872_radiative} reported the surprising observation that the branching fraction for $Y(4260) \rightarrow X(3872) \gamma$
  is a factor $\simeq$50 higher than that of an E1 charmonium transition. This particular decay also represents the unique observation of an 
  electromagnetic transition between two a-priori-different classes of exotic states, namely $X$ and $Y$ states, and thus might indicate a common nature. 
  As mentioned above, Belle II has improved capabilities for photon detection with the upgraded electromagnetic calorimeter.
  Radiative decays can be detected from very low photon energies of 20~MeV, inaccessible by many other experiments, up to $\geq$4~GeV. 
  The envisaged high statistics will compensate the EM suppression of $\simeq$(1/137)$^2$ for the branching fraction of these decays, 
  and enables searches for more transitions among exotic states, or for the first time between exotic states and conventional states.}

\item Absolute branching fractions are important, as one single measurement may impact dozens of entries in the PDG. At $B$ factories, where the $\Upsilon(4S)$ decays at rest to the $B \overline{B}$ meson pair, the measurement of those is possible. All branching fractions of a given particle in the final state of the $B$ meson decays can be normalised this way. The technique uses {\it (a)} $e.g.$ the recoil mass to a particle under investigation in a two-body decay of the $B$ meson, which is reconstructed as signal \cite{belle_x3872_absolute}, 
  and {\it (b)} tagging the opposite-side $B$ meson, for the normalisation of the absolute yield.
  BaBar succeeded to measure the absolute branching fraction of $B \rightarrow K X(3872)$, equal to (2.1 $\pm$ 0.7) $\times$ 10$^{-4}$ \cite{babar_x3872_absolute}, with the above-mentioned large impact, as this single measurement defines 100\% of all $X$(3872) decays. 
Belle II will be able to reduce the {\color{blue} statistical} error to $\leq$0.2 ($\leq$0.1) with 10~ab$^{-1}$ (50~ab$^{-1}$). 
  
\end{itemize}

\vspace*{-0.8cm}

%%%%%%%%%%%%%%%%%%%%
\subsection{Baryons}
%%%%%%%%%%%%%%%%%%%%

\vspace*{-0.3cm}

%%%%%%%%%%%%%%%%%%%%%%%%%%%%%%%
\subsubsection{Strange baryons}
%%%%%%%%%%%%%%%%%%%%%%%%%%%%%%%

\vspace*{-0.3cm}

Baryons can be produced at Belle II in the $e^+ e^- \rightarrow$ baryon antibaryon process via ISR or in the continuum, associated with other hadrons,
as well as in the B-mesons decays. Strange baryons are of particular interest because of their possibly exotic internal structure beyond the ordinary three-quark configuration \cite{hyodo}.
An example is the evidence for the two-pole structure in the $\Lambda$(1405) \cite{pdg}. 
As cross sections are estimated in the range of 10$-$100~pb, 
with an assumed reconstruction efficiency of $\simeq$20\% \cite{sumihama}, 
we expect $10^8$$-$$10^9$ events with excited $\Lambda$, $\Sigma$ or $\Xi$ baryons in a data set of 50~ab$^{-1}$.
In addition, baryons can also be produced in $\Upsilon(nS)$ decays.
At small momenta, the production of those is enhanced  a factor $\simeq$2
with respect to production in the continuum \cite{cleo_strange_baryon}.

\vspace*{-0.3cm}

%%%%%%%%%%%%%%%%%%%%%%%%%%%%%%%
\subsubsection{Charmed baryons}
%%%%%%%%%%%%%%%%%%%%%%%%%%%%%%%

\vspace*{-0.3cm}

After decades of searches, the $\Xi_{cc}^{++}$ as the first double charmed baryon was discovered by LHCb \cite{lhcb_xicc++}. 
One of the recent hot topics is the search for doubly charmed $\Xi_{cc}^+$, 
the isopin partner of the $\Xi_{cc}^{++}$, which has not been yet observed.
The masses of these states are important input for predictions of binding energies for stable tetraquarks ($cc\overline{q}\overline{q}$)
by heavy quark symmetry \cite{Tcc_quigg} \cite{Tcc_rosner}. 
$\Xi_{cc}$ states can be searched in their decays to $\Xi_c$ states. At Belle, the $\Xi_c^0$ and $\Xi_c^+$ were reconstructed in 10 and 7 different exclusive decay channels, respectively. 
At Belle II, with a data set of 50~ab$^{-1}$, 5.0$\times$10$^6$ and 2.8$\times$10$^6$ reconstructed decays
are a realistic extrapolation based upon Belle numbers \cite{belle_xic0_xic+}, respectively.
In addition, the search for new $\Xi_c$ states remains an important topic. Belle observed two orbitally excited $\Xi_c^0$ states 
in their radiative decays \cite{belle2_xic0_orbital}. 
For the $\Xi_c^+$ an absolute branching fraction (see above) was measured by Belle \cite{belle_xic+_absolute}.
The surprising observation was that only 24.4\% of the decays are known and listed in the PDG.
{\color{soeren} As examples, Belle found $\Xi_c$ decays with internal $W$ exchange \cite{belle_xic_W_exchange},
semileptonic decays \cite{belle_xic_semileptonic} and Cabbibo-suppressed decays \cite{belle_xic+_absolute}. 
Belle II is expected to contribute significantly to other searches, not only for $\Xi_c$, but also for $\Sigma_c$ states,
for which so far less states have been observed compared to $e.g.$ excited $\Lambda_c$ states.
Furthermore, the absolute branching fraction of the $\Omega_c$ has not been measured yet. 
In general, $J^{PC}$ values for many charmed baryons are unknown yet, and can be determined by Belle II using angular distributions.}

Lifetimes of charmed baryons can also be measured at Belle II with high precision. 
Recent results are summarized in Tab.~\ref{tlifetime}, in comparison to measured lifetimes of charmed mesons.
Belle II confirmed the surprising observation by LHCb, that, contrary to the expectation by HQET,
the lifetime $\Omega_c^0$ is smaller than that of the $\Lambda_c^+$.
For the charmed mesons, these are the world’s most precise measurements
with accuracies of 3.5 and 5.4 per mill, for $D^0$ and $D^+$, respectively. 

\begin{table}[hhh]
  \begin{center}
  \begin{tabular}{|l|l|l|}
    \hline
    Particle & Measured lifetime & Reference\ \\
    \hline
    \hline
    $\Lambda_c^+$ & 203.2 $\pm$ 0.9 $\pm$ 0.8~fs & \cite{belle2_lifetime_lambdac+} \\
    $\Omega_c^0$ & 243 $\pm$ 48 $\pm$ 11~fs & \cite{belle2_lifetime_omegac+} \\
    \hline
    $D^0$ & 410.5 $\pm$ 1.1 $\pm$ 0.8~fs & \cite{belle2_lifetime_D0_D+} \\
    $D^+$ & 1030.4 $\pm$ 4.7 $\pm$ 3.1~fs & \cite{belle2_lifetime_D0_D+} \\
    \hline
  \end{tabular}
  \end{center}
  \caption{Measurements of lifetimes of charmed baryons and charmed mesons at Belle II.\label{tlifetime}}
  \label{tlifetime}
\end{table}

\vspace*{-0.8cm}

%%%%%%%%%%%%%%%%%%%%%%%%%%%%%%%%%%%%%%%%%%%
\subsubsection{Dibaryons and Antideuterons}
%%%%%%%%%%%%%%%%%%%%%%%%%%%%%%%%%%%%%%%%%%%

\vspace*{-0.3cm}

The loosely bound $H$ dibaryon was predicted by Jaffe \cite{jaffe_dibaryon}
and has been searched for by many experiments, among them Belle \cite{belle_dibaryon}.
{\color{blue} Due to its specific hexaquark configuration ($uuddss$) with three different flavors,
color-hyperfine interaction could provide the binding.}
Belle II offers the unique possibility for searches in bottomonium decays such as
$\Upsilon(3S) \rightarrow \overline{\Lambda} \overline{\Lambda} H$ + hadrons.
This requires changing the c.m.\ energy to the $\Upsilon(3S)$,
which is below the $B \overline{B}$ threshold and planned as part of the Belle II physics program. 
A extrapolation of the expected yield in a data set of 300~fb$^{-1}$ Y(3S)
is about 60 Million events with one $\Lambda$ or $\overline{\Lambda}$,
and about 3 Million events with one $\Lambda$ $\overline{\Lambda}$ pair \cite{bianca_hadron21}.
The projected upper limit estimate is less than 5$\times$10$^{-7}$ for an assumed mass of 1.179~$GeV/c^2$.  

In addition, antideuterons have been observed in $\Upsilon(3S)$ decays \cite{argus_antideuteron} \cite{cleo_antideuteron}.
Interestingly, an analysis by BaBar \cite{babar_antideuteron} showed that the antideuterons are partially produced from gluon-dominated $\Upsilon$ decays
and partially from $e^+ e^- \rightarrow q \overline{q}$.
The momentum distributions is in good agreement with expectation from coalescense, $i.e.$ the overlap of the nucleon wave functions in the final state.
In fact, they can be described well with a fireball model, which is quite surprising for an $e^+ e^-$ collision. 
With a data set of 300~fb$^{-1}$ $\Upsilon$(3S) decays one would collect about 3$\times$10$^4$ antideuterons. 
This would enable the world’s best estimate of coalescence parameter, which could be useful input to nuclear experiments.
The data would also enable search for possibly excited deuteron states, such as the $d^{*}$ resonance observed at WASA-at-COSY \cite{d_cosy},
which represents another possible hexaquark candidate. 

\vspace*{-0.3cm}

%%%%%%%%%%%%%%%%%%%%%%%%%
\section{Inputs to $g-2$}
%%%%%%%%%%%%%%%%%%%%%%%%%

\vspace*{-0.3cm}

As mentioned in the LRP 2017, the determination of the muon $g-2$ anomaly, often defined as $a_{\mu} = (g-2)/2$, is to be determined with high precision, as it offers sensitive impact to physics beyond the Standard Model (SM).  The largest contribution to the uncertainty on the SM value of $a_{\mu}$ is from the hadronic vacuum polarization (HVP) process.  The contribution from HVP is evaluated by performing a dispersion integral over the cross section for $e^+e^- \rightarrow$ hadrons. This calculation contributes approximately 67\% of the total theoretical uncertainty on  $a_{\mu}$. The largest contribution to the HVP integral is from the proces $e^+e^-\to \pi^+\pi^-$,  which has been measured by several experiments at different c.m. energies. The systematic uncertainty in previous measurements of the cross section is reported as 0.9\% by BES III \cite{bes3_pipi}, 0.7\% by KLOE \cite{kloe_pipi}, and 0.5\% by BaBar \cite{babar_pipi}. However, the latter measurement is limited to the dipion invariant mass range 0.6-0.9~GeV/$c^2$ [23].
{\color{soeren} Furthermore, there is 2.9$\sigma$ disagreement between the BaBar and KLOE measurements, which dominates the HVP-related uncertainty on $a_{\mu}$.} The goal of Belle II is a 0.5\% systematic uncertainty in the dipion invariant mass range up to 2~GeV/$c^2$. This measurement will be performed on the data sample already collected by Belle II. {\color{soeren} At present the new energy scan measurements are ongoing at VEPP-2000, aiming at a precision of 0.5-0.3\%,
  and the accurate comparison of two alternative approaches is important.}

The second largest contribution of $10\%$ is from the process $e^+e^-\to\pi^+\pi^-\pi^0$ \cite{sugBaBar}. Belle II is targeting a precise cross-section measurement with a systematic uncertainty smaller than 2\%. The measurement exploits Belle II's excellent capabilities for reconstructing photons and $\pi^0$ mesons.
 
Belle II also provides important input to the interpretation of the $(g-2)_\mu$ measurements in terms of dark-sector particles. A search for $Z^{\prime}$ in the process $e^+e^-\to \mu^+\mu^-Z^{\prime}$, where the $Z^{\prime}$ decays to undetectable particles, is able to exclude such a $Z^{\prime}$ explanation of the $(g - 2)_{\mu}$ anomaly in the range of $Z^{\prime}$ mass from 0.8 to $5.0~\mathrm{GeV}/c^2$. This measurement uses a data set corresponding to $79.7~\mathrm{fb^{-1}}$ \cite{belle2_zprime}. Therefore, with a $5~\mathrm{ab}^{-1}$ ($50~\mathrm{ab}^{-1}$) data sample, Belle II will reach sensitivity to masses as small as 0.2 (0.04)~$\mathrm{GeV}/c^2$ \cite{snowmass_zprime_reach}.

\vspace*{-0.6cm}

%%%%%%%%%%%%%%%%%%%%%%%%%%%%%
\section{Proton form factors}
%%%%%%%%%%%%%%%%%%%%%%%%%%%%%

\vspace*{-0.3cm}

{\color{blue} The proton form factor can be measured in $e^+ e^- \rightarrow p \overline{p}$.
The effective formfactor, which contains terms of the electric and the magnetic formfactors, $G_E$ and $G_M$, respectively, 
was measured by BaBar \cite{babar_proton_effective_formfactor} up to invariant masses $m_{p \overline{p}}$$\simeq$4.4~GeV/$c^2$,
which is about 0.2~GeV/$c^2$ more than at any other experiment.
The magnetic form factor was measured up to $m_{p \overline{p}}$=6~GeV/$c^2$,
representing the only measurement worldwide in a regime $|$ $G_M$ $|$$<$10$^{-3}$.  
This is a very interesting region, as above $m_{p \overline{p}}$=5~GeV/$c^2$,
data seem to indicate that timelike (from $e^+ e^-$ collisions) and spacelike (from $e p$ collisions) form factors are asymptotically converging. 
The measurements are statistics limited and Belle II will be able to extend the search even to higher masses.}

\vspace*{-0.6cm}

%%%%%%%%%%%%%%%%%%%%%%%%%%%%%%%%%
\section{Fragmentation functions}
%%%%%%%%%%%%%%%%%%%%%%%%%%%%%%%%%

\vspace*{-0.3cm}

{\color{red} Understanding hadronization is one of the QCD frontiers, as the process eludes perturbative calculations. 
Belle II provides unique opportunities to elucidate hadronization dynamics by measuring correlations in $e^+ e^-$ annihilations. 
Those can be used for the precise determination of fragmentation functions (FFs), describing the fragmentation of light quarks into hadrons.
Since FFs are non-perturbative objects, they have to be measured in experiments.
They can be seen as the time-like counterparts to parton distribution functions (PDFs), But unlike PDFs they are presently inaccessible by lattice QCD. FFs are crucial ingredients to extract the spin structure of the nucleon.
For example, the first extraction of the so-called Collins fragmentation function at Belle
enabled the extraction of transversity PDF \cite{collins1} \cite{collins2},
one of the three collinear PDFs needed to describe the nucleon.  
Belle also observed, for the first time, the transverse Lambda polarizing FF,
which is sensitive to the gauge structure of QCD \cite{belle_lambda_spin} \cite{boer_lambda_spin}.
Therefore, FFs are important input to measurements 
at next generation SIDIS (semi-inclusive deep inelastic scattering) experimentes (e.g.\ at JLab12),
$pp$ experiments (e.g.\ COMPASS)
and at the future electron-ion collider EIC. 
For details, see \cite{snowmass_qcd}.

Belle II will be able to measure di-hadron FFs, which, contrary to inclusive single-hadron FFs,
capture the dynamics in the hadronization process and provide sensitivity not only to the transverse,
but also to londitudinal quark momentum \cite{snowmass_qcd}.
The Belle II data will also enable precision tests of hadronization models as implemented in MC generators (e.g.\ Pythia, SHERPA, HERWIG), 
and verify the energy dependence of these models in conjunction with LEP data. 
For the first time the large statistics will allow comparisons with novel polarized MC hadronization models \cite{snowmass_qcd}.}

\vspace*{-0.6cm}

%%%%%%%%%%%%%%%%%%%%%%%%%%%%%%%%
\section{Hadronic $\tau$ decays}
%%%%%%%%%%%%%%%%%%%%%%%%%%%%%%%%

\vspace*{-0.3cm}

The weak currents coupled to $W^\pm$ can be classified in terms of parity and $G$-parity as follows:
first-class currents with $J^{PG}$ = $0^{--}$, $1^{-+}$, $1^{+-}$; second-class currents (SCC) with $J^{PG}$ = $0^{+-}$, $1^{++}$.
In the SM with isospin (hence $G$-parity) conservation only first-class currents exist.
Still SCC can be induced in decays to hadrons composed of light quarks by isospin violation, $i.e.$ the charge and mass differences between up and down quarks.
Corresponding hadronic $\tau$ decays have not been observed yet.
The discovery and the measurement of the branching fraction is important,
because it would provide a testing ground for contributions to the SCC from genuine effects beyond the standard model. 
An example is the decay $\tau^{\pm} \rightarrow \pi^{\pm} \eta \nu_{\tau}$, for which the branching fraction is expected to be small,
as low as 10$^{-5}$ \cite{tau_pich} or less.
Measured upper limits are 9.9$\times$10$^{-5}$ \cite{tau_babar} and 7.3$\times$10$^{-5}$ \cite{tau_belle}, by BaBar with 470~fb$^{-1}$ and Belle with 670~fb$^{-1}$, respectively.  

Belle II will further improve these measuements: an upper limit of 1$\times$10$^{-5}$ is extrapolated for 20~ab$^{-1}$.

%\vspace*{-0.3cm}
\newpage

%%%%%%%%%%%%%%%%%%%%%%%%%%%%%%%%%%%%%%%%%%%%%
\section{Strong coupling constant $\alpha_S$}
%%%%%%%%%%%%%%%%%%%%%%%%%%%%%%%%%%%%%%%%%%%%%

\vspace*{-0.3cm}

{\color{blue} The present value of the strong coupling constant at the reference $Z$ pole mass amounts to $\alpha_S$($m_Z^2$)=0.1179$\pm$0.0009 \cite{pdg}.  Thus, the uncertainty $\delta$$\alpha_S$/$\alpha_S$$\simeq$0.8\% is orders of magnitude larger than that of the other three fundamental interactions (QED, weak, and gravitational) couplings. Improving our knowledge of $\alpha_S$ is crucial, in particular to reduce the theoretical uncertainties in the
  calculations of perturbative QCD observables, such as cross sections, decay rates or masses.
  
Belle II can determine $\alpha_S$ in hadronic $\tau$ decays, by measuring the ratio $R(s) = \sigma(e^+ e^- \rightarrow hadrons; s) / \sigma(e^+ e^- \rightarrow \mu^+ \mu^- ; s)$
in a range of the c.m.\ energy $\sqrt{s}$=2-10~GeV, and using $e^+e^-$ event shapes. For details, see \cite{snowmass_qcd}.}

\end{document}